\documentstyle[aps,prb,eqsecnum,multicol,epsf]{revtex}
\begin{document}
\input epsf
\bibliographystyle{prsty}
\draft
\preprint{NSF-TIP}
\title{Criticality in the two-dimensional random-bond Ising model}
\author{Sora Cho}
\address{Department of Physics, University of California,
        Santa Barbara, California 93106}
\author{Matthew P.A. Fisher}
\address{Institute for Theoretical Physics,
        University of California, Santa Barbara, California 93106\\
	and Department of Physics, University of California,
        Santa Barbara, California 93106}
\date{\today}
\maketitle

\begin{abstract}
The two-dimensional (2D) random-bond Ising model has a novel
multicritical point on the ferromagnetic to paramagnetic phase boundary.
This random phase transition is one of the simplest examples
of a 2D critical point occurring at both finite temperatures
and disorder strength.  We study the associated critical properties,
by mapping the random 2D Ising model onto a network model.  The model 
closely resembles network models of quantum Hall plateau transitions, 
but has different symmetries.  Numerical transfer matrix calculations
enable us to obtain estimates for the critical exponents at the random 
Ising phase transition.  The values are consistent with recent estimates
obtained from high-temperature series.

\end{abstract}

\begin{multicols}{2}
\section{Introduction}
\label{sec:intro}
Two-dimensional (2D) models have played a special role in the theory 
of phase transitions.\cite{sft89}
In 1944 Onsager's exact solution of the 2D Ising model 
gave critical
exponents that were simple rational numbers,
although different than Landau theory.
In the 1970 renormalization group (RG) calculations
revealed exponents varying continuously below
an upper critical dimension, illustrating the breakdown
of Landau theory.  
But it was unclear why the 2D Ising exponents and those
for other exactly soluble 2D models were rational numbers.
This fact was explained by the remarkable development of
conformal field theory in the 1980s.  Under the assumption
of conformal invariance at criticality, 
it was possible to analyze 
a large
class of 2D critical points.\cite{sft89}
Moreover, a first step was made towards a full classification
of {\it all} allowed 2D phase transitions.

Many physically important 2D phase transitions occur in systems
with quenched disorder.  An example of particular
experimental interest is the transition between
plateaus in the integer quantum Hall effect (IQHE).\cite{huckestein95} 
This transition has been successfully
studied numerically, but so far has eluded
analytic treatments via either RG calculations, exact methods or
conformal field theory.  Given the general power
of conformal field theory in 2D, it has been
surprisingly unhelpful in understanding such random phase transitions.

In this paper we analyze a nontrivial 2D random phase transition
which occurs in the simplest of all models: 
the 2D Ising model with random bonds.  
Our approach is numerical, and closely parallels 
earlier work on the IQHE transition.\cite{qhe90}
We first map the 2D random Ising model into a variant
of the Chalker-Coddington\cite{chalker88} network model, which describes
noninteracting chiral fermions.  The random 
Ising transition corresponds to a fermion localization transition.
We extract critical exponents numerically by standard transfer matrix methods.  The values are
in reasonable agreement with those recently obtained 
by Singh and Adler\cite{singh96}
via high-temperature series.  Unfortunately,
this random Ising transition has also eluded any analytic
treatment.

To be more specific, consider the Ising model on a 2D square lattice,
\begin{equation}
H_J = - \sum_{\langle i,j\rangle} J_{ij} S_iS_j,
\end{equation}
with nearest-neighbor interactions $J_{ij}$ taken as random variables
with a distribution $P_J(J_{ij})$.  
For the simple distribution
$P_J(J_{ij}) = p\delta(J_{ij}+J) + (1-p)\delta(J_{ij}-J)$,
corresponding to a fraction $p$ of antiferromagnetic
bonds, the phase diagram, as established by various
methods,\cite{binder86,young77,maynard82,ozeki87,mcmillan86} 
is shown schematically in Fig. 1.  

\begin{figure}
\epsfxsize = 2.5in
\centerline{\epsfbox{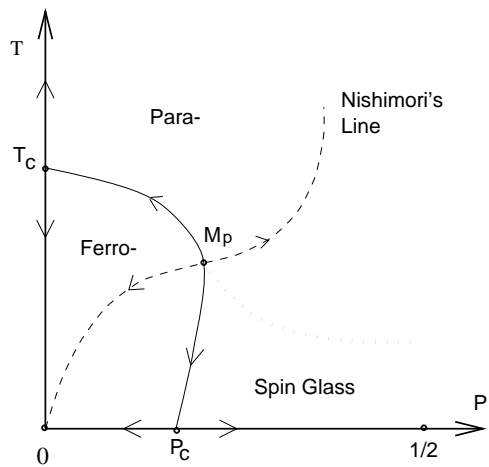}}
\begin{minipage}[t]{8.1cm}
\caption{ Schematic phase diagram of the 2D $\pm J$
random-bond Ising model.}
\end{minipage}
\end{figure}

For small $p$ there
is a phase boundary separating the ferromagnetically ordered
phase at low temperatures from the paramagnet.  A larger $p$ destroys the
ferromagnetic phase, replacing it in high dimensions ($d \ge 3$)
by a spin glass phase.  In this case a multicritical point is expected
at the coexistence point of all three phases.  In 2D the spin glass phase
is not present at $T \ne 0$, being destroyed by 
thermal fluctuations.\cite{binder86}
However, 
the multicritical point still exists on the ferro-to-para phase
boundary.\cite{ueno91,kitatani90,kitatani92,ozeki90}  The multicritical
point is unstable along the phase boundary, with RG flows
as depicted in Fig. 1, consistent with the
(marginal) irrelevance of weak disorder
at the pure 2D Ising critical point ($p=0$).  LeDoussal and Harris
\cite{ledoussal88}
have argued that the Nishimori line, along which the internal 
energy is analytic,\cite{nishimori81}
passes through the multicritical point and coincides with one of the
two RG scaling axes.  The other scaling axis
is tangent to the ferro-to-para phase boundary.  Recently,
Singh and Adler\cite{singh96} have obtained estimates for the two associated
critical exponents, using a high-temperature series method.
 From a general point of view, this multicritical point is of interest,
being probably the simplest 2D critical point which occurs at
both finite temperature and finite disorder strength.  

Our paper is organized as follows.  In Secs. II and III,
we show that the random-bond 2D Ising model can be mapped, using a fermionic
representation, to a variant of the Chalker-Coddington 
network model.\cite{chalker88}
This mapping reveals a close similarity between 
the random Ising transition and the IQHE plateau transition.
However, due to a symmetry difference, the two transitions are {\it not} in the
same universality class.
In Sec. IV, we employ a transfer matrix approach to 
analyze the network model, and obtain estimates for the exponents
at the random Ising multicritical point.
Sec. V gives a brief summary and conclusion.  
 
\section{Fermionic Representation}
\label{sec:diag}

It is well known that the critical properties of the
pure 2D Ising transition are equivalent to a massless
Majorana (real) fermion field.\cite{sft89}  
In his studies of the bond-diluted
Ising model,
Shankar\cite{shankar87} constructed a model in terms of conventional
(Dirac) fermions, by adjoining two identical copies. 
For the random-bond Ising model, we show below
that this procedure leads to a model
of 2D chiral fermions, with a hopping matrix element of random {\it sign}.

Following Shankar,\cite{shankar87} we
consider a spatially anisotropic Ising model, retaining a lattice in
one direction, but taking the continuum limit in the other 
(the ``imaginary time" direction).
The partition function, when expressed in terms of a transfer
matrix, can then be written $Z=$Tr $\exp(-\beta H_{1D})$, where
$H_{1D}$ is a 1D quantum Hamiltonian 
and $\beta$ is the system size in the ``time"
direction.  The appropriate 1D Hamiltonian for the pure Ising model is,
\begin{equation}
H_{1D} = \sum_{n}[\kappa_1 \sigma_{n}^x +
\kappa_2 \sigma_{n}^z\sigma_{n+1}^z] ,
\end{equation}
where $n$ label sites of a 1D lattice and $\sigma^\alpha$ with $\alpha=x,y,z$
are Pauli matrices.  This model
exhibits a phase transition when $\kappa_1 = \kappa_2$,
which is in the (pure) Ising universality class, as verified
below.  The transition also follows from a duality symmetry which exchanges
high- and low-temperature phases.\cite{sft89}  With the definition
$\sigma_n^x \equiv \mu_n^z\mu_{n+1}^z$ and $\sigma_n^z \equiv \prod_{m<n}
\mu_m^x $, the Hamiltonian can be written in the form (2.1) with
$\sigma^\alpha \rightarrow \mu^\alpha$ and 
$\kappa_1 \leftrightarrow \kappa_2$. 

The partition function is also invariant under $\kappa_i \rightarrow
- \kappa_i$ ($i=1,2$), as seen by a spin rotation,
$\sigma_n^\alpha \rightarrow - \sigma_n^\alpha$, with $\alpha = x,y$ for $n$ odd
and $\alpha=x,z$ for $n$ even, which restores the Hamiltonian
to its original form.  This transformation is equivalent
to $J \rightarrow - J$ in in Eq. (1.1), mapping from a ferromagnetic to
antiferromagnetic spin model.   Thus randomness
in the sign of the exchange interaction in Eq. (1.1) is equivalent
to randomness in the {\it sign} of $\kappa_i$.  
To describe the
$\pm J$ Ising model, we thus consider interactions
which are random functions of both $n$ and imaginary time $\tau$,
$\kappa_i \rightarrow \kappa_i(n,\tau)$, taking either sign. 

A fermionic representation can be obtained by introducing the Majorana fields
\begin{equation}
\eta_1(n) \equiv {1 \over \sqrt{2}} \prod_{m<n} \sigma_m^x\sigma_n^y,
\hspace{.1in}
\eta_2(n) \equiv {1 \over \sqrt{2}} \prod_{m<n} \sigma_m^x\sigma_n^z,
\end{equation}
which anticommute, $\{\eta_i(n),\eta_j(m)\}=\delta_{ij}\delta_{nm}$.
The Hamiltonian is quadratic when expressed in terms of these new variables,
\begin{equation}
H_{1D} = (-2i)\sum_{n} [\kappa_1\eta_1(n)\eta_2(n) -
\kappa_2\eta_1(n)\eta_2(n+1)].
\end{equation}

An identical copy of the system is constructed 
by introducing a new set of Majorana fields $\xi_i$
The Hamiltonian obtained by summing
the two, 
$\tilde{H} \equiv {1\over 2} [H_{1D}(\eta_i) + H_{1D}(\xi_i)]$,
can be
expressed in terms of standard
(Dirac) fermion operators
\begin{equation}
\psi_i \equiv {1 \over \sqrt{2}} (\eta_i + i\xi_i), \hspace{.1in}
\psi_i^\dagger \equiv {1 \over \sqrt{2}} (\eta_i - i\xi_i)
\end{equation}
as 
\begin{eqnarray}
\tilde{H} & = & \sum_{n}(-i\kappa_1)[\psi_1^\dagger(n)\psi_2(n) 
- \psi_2^\dagger(n)\psi_1(n)]
\nonumber \\
    &   & + (i\kappa_2)[\psi_1^\dagger(n)\psi_2(n+1) - \psi_2^\dagger(n+1)\psi_1(n)]. 
\end{eqnarray}
Notice that $\tilde{H}$ 
has a conserved U(1) charge:
$\psi_1^\dagger\psi_1+\psi_2^\dagger\psi_2$.  At the pure
Ising transition, there are gapless excitations in this conserved charge.
With disorder present, the transition corresponds to
a localization transition of these conserved fermions.

To complete the mapping, we
express the partition function as a path integral over Grassmann
fields,
\begin{eqnarray}
Z=\int {\cal D}\psi {\cal D} \bar{\psi} \exp(-S),
\end{eqnarray}
where $S$ is the Euclidean Action for $\tilde{H}$:
\begin{equation}
S = \int_\tau \sum_{n} [\bar{\psi_1} (n)\partial_{\tau}\psi_1(n)
 + \bar{\psi_2} (n)\partial_{\tau}\psi_2(n)] + \tilde{H}(\bar{\psi},\psi).
\end{equation}
Reinterpreting imaginary time as a spatial
coordinate, $\tau \rightarrow x$, the action $S$ can be viewed
as a 2D Hamiltonian of chiral fermions, denoted ${H}_{2D}$.
To bring it into a canonical form, we define new right- and left-moving
fermion fields
\begin{equation}
\psi_{Rn} = (-1)^n \psi_1(n),\hspace{0.1in}   \psi_{Ln} = (-1)^n \psi_2(n),
\end{equation}
\begin{equation}
\psi^\dagger_{Rn} = i (-1)^n \bar{\psi}_1(n),\hspace{0.1in} 
\psi^\dagger_{Ln} = -i (-1)^n \bar{\psi}_2(n).
\end{equation}
In terms of these the action becomes
\begin{eqnarray}
\it{H}_{2D} & = & \int dx\sum_{n} [\psi^\dagger_{Rn}(i\partial_x)\psi_{Rn} +
\psi^\dagger_{Ln}(-i\partial_x)\psi_{Ln} \nonumber\\ &  & + \kappa_1
(\psi^\dagger_{Rn}\psi_{Ln} + \psi^\dagger_{Ln}\psi_{Rn})\nonumber\\ 
& & + \kappa_2(\psi^\dagger_{Rn}\psi_{Ln+1} +\psi^\dagger_{Ln+1}\psi_{Rn})].
\end{eqnarray}

This Hamiltonian has a simple pictorial 
representation in terms of 1D right- and left-moving fermion fields,
coupled together by hopping 
strengths $\kappa_1$ and $\kappa_2$, as depicted in Fig. 2.
The model closely resembles an anisotropic version of
the Chalker-Coddington network model.\cite{chalker88}   In the next section
we describe a lattice version, appropriate
for numerical simulations.

\begin{figure}
\epsfxsize = 2.5in
\centerline{ \epsfbox{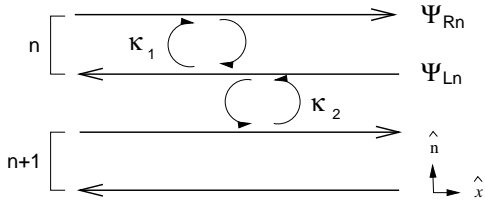}}
\begin{minipage}[t]{8.1cm}
\caption{ pictorial representation of the chiral
fermion Hamiltonian (2.10).  Neighboring chiral modes, propagating in the
direction of the arrows, are coupled via tunneling matrix elements $\kappa_i$.}
\end{minipage}
\end{figure}

In the absence of disorder, with 
$\kappa_1$ and $\kappa_2$ constant, $H_{2D}$ can be easily diagonalized
by transforming to momentum space.
The energy eigenvalues
satisfy
\begin{equation}
E^2=p_x^2+\kappa_1^2+\kappa_2^2 + 2\kappa_1\kappa_2 \cos p  ,
\end{equation}
with $p_x$ the x component of momentum
and $p$ 
a transverse momentum in the range $-\pi$ to $\pi$.
The energy is a minimum when $p=\pi$ and $p_x = 0$,
and given by $|E|_{\rm min} = \pm |\Delta|$, with $\Delta = \kappa_1 - \kappa_2$.
The pure Ising critical point occurs when $\Delta=0$.
For $\Delta \ne 0$, there are no zero energy eigenvalues 
of the 2D Hamiltonian.  If a wave of energy $E=0$ is incident
in the x direction, it will decay as $\exp(-|\Delta| x)$,
since $p_x = i \Delta$ is pure imaginary.  The decay length
$\xi \sim |\Delta|^{-1}$ corresponds to the correlation length of the pure 2D Ising model.  The critical exponent is $\nu_{pure} = 1$ as expected.  

With disorder, the tunneling amplitudes, $\kappa_i$,
become random functions of position, $n$ and $x$, and
momentum is not a good quantum number.
Nevertheless, at the Ising multicritical point
(see Fig. 1), one expects the $E=0$ states of $H_{2D}$
to be {\it extended}, corresponding to an infinite correlation length.
Away from criticality, 
one anticipates {\it localized} electronic states at $E=0$,
rather than a gap as in the pure case.

\section{Network model}

In pioneering work, Chalker and Coddington\cite{chalker88} introduced a network
model to study numerically the transition between
IQHE plateaus.  This model is essentially a lattice version
of a chiral fermion Hamiltonian, similar to Eq. (2.10).
The model consists of links and nodes, as depicted in Fig. 3.
On each link there is a complex amplitude representing
the fermion (electron) wave function.  At the nodes,
two incoming wave functions scatter into two outgoing ones,
conserving probability.  The nodes are specified by an S matrix.
In the original network model,\cite{chalker88} the complex amplitudes
acquired a random phase factor upon propagating along
a given link, corresponding physically to a (random)
magnetic flux through plaquettes.   

\begin{figure}
\epsfxsize = 2.7in
\centerline{ \epsfbox{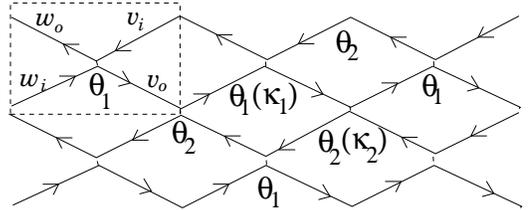}}
\begin{minipage}[t]{8.1cm}
\caption{ Representation of network model with width $L=4$ and length
$N=3$.  The arrows indicate the direction of wave propagation
along links.  The parameters $\theta_i$ specify scattering at the nodes.}
\end{minipage}
\end{figure}

On physical grounds, it is clear that a very similar network
model should suffice for describing the propagation of $E=0$
waves of the Hamiltonian (2.10).  In the appropriate network model,
the tunneling amplitudes $\kappa_i$ are replaced by node
parameters.  Since we are interested in $E=0$,
there are $\it no$ phase factors associated with the links themselves.
The network model is specified by a transfer matrix
$T$, taken, say, in the horizontal direction in Fig. 3.
This matrix is decomposed into a product of matrices $M_j$,
representing columns of the network, with $j$ running from $1$ to $N$,
the length of the network.
Each matrix $M_j$ is a product of two matrices $M_j = A_j(\theta_1) B_j(\theta_2)$, representing
two adjacent nodes in Fig. 3.  The two node parameters
$\theta_1$ and $\theta_2$ correspond to the two hopping
coefficients $\kappa_1$ and $\kappa_2$.

The matrix representing tunneling at a given node 
is constructed to conserve the current or, equivalently, the
U(1) charge.
Following Chalker and Coddington,\cite{chalker88} the node
in the dotted box in Fig. 3 is written as 
\[  \left( \begin{array}{c} w_{\rm out} \\ 
w_{\rm in} \end{array} \right) =
 \left( \begin{array}{cc}
    \cosh\theta_1 & \sinh\theta_1 \\ \sinh\theta_1 & \cosh\theta_1
\end{array} \right) 
\left( \begin{array}{c} v_{\rm in} \\ v_{\rm out} \end{array} \right). \] 
By construction, this matrix conserves the current,
$|w_{\rm in}|^2+|v_{\rm in}|^2 =|w_{\rm out}|^2+|v_{\rm out}|^2$.
Moreover, an incident wave, say, $w_{\rm in}$, is backscattered into $w_{\rm out}$
with probability $\tanh^2(\theta_1)$.  Since this tunneling
probability is proportional to $\kappa_1^2$ in the 2D Hamiltonian (2.10),
we make the identification
\begin{equation}
\tanh(\theta_i) \leftrightarrow \kappa_i  .
\end{equation}
Although the Hamiltonian (2.10) is intrinsically anisotropic,
the network model can be made invariant\cite{chalker88} under a $\pi/2$ spatial rotation
by choosing
$\sinh(\theta_1) \sinh(\theta_2) =1$, at successive nodes.  

It is instructive to briefly consider the pure network model, with
constant node parameters.  In this case, the transfer matrix can be
diagonalized in momentum space.  
For a network with length $N$
and width $L$, the total transfer matrix is $T=M_0^N$, with $M_0$ an $L\times L$ matrix.  Due to translational invariance, the eigenvalues of $M_0$, and hence
$T$, can be labeled by a transverse momentum $p$.  We denote these
as $\lambda(p)$.  For a given transverse momentum $p$,
these follow as eigenvalues of a simple 2 by 2 matrix:    
\end{multicols}
\[ \left( \begin{array}{ll}
\cosh\theta_1 \cosh\theta_2 + \sinh\theta_1 \sinh\theta_2
e^{ip} & \sinh\theta_1 \cosh\theta_2+ \cosh\theta_1 
\sinh\theta_2 e^{ip} \\
\sinh\theta_1 \cosh\theta_2 + \cosh\theta_1 \sinh\theta_2
e^{-ip} & \cosh\theta_1 \cosh\theta_2 + \sinh\theta_1
\sinh\theta_2 e^{-ip}  
\end{array} \right). \]
\begin{multicols}{2}
 
One finds,
\begin{equation}
\lambda_{\pm}(p) =  A \pm \sqrt{A^2 -1}  ,
\end{equation}
with\begin{equation}
A = \cosh(\theta_1 - \theta_2) + \sinh(\theta_1) 
\sinh(\theta_2) [1 + \cos(p)]  .
\end{equation}
As before, at the pure Ising critical point a nondecaying mode
is expected.
Since the eigenvalues of the total transfer matrix are $\lambda^N$,
this is only possible if $|\lambda|^2 =1$.  This requires $p =\pi$ and
$\theta_1 = \theta_2$, 
the expected condition for Ising criticality.  Specializing to $p = \pi$
and the isotropic case, with $\sinh(\theta_1)\sinh(\theta_2)=1$,
the eigenvalues take the simple form
\begin{equation}
\lambda_+ = {1\over\lambda_-}={1+\Delta\over 1-\Delta},
\end{equation}
with $\Delta$ measuring the ``distance" to the critical point:
\begin{equation}
\Delta = \tanh(\theta_1) - \tanh(\theta_2)  .
\end{equation}
These eigenvalues describe the slowest decay of $T \sim \lambda^N$.
The Ising correlation length follows as $\xi = 1/ln(\lambda_+)$,
and as expected varies as $\xi \sim 1/|\Delta|$ upon approaching
the critical point, $\Delta \rightarrow 0$.   

We now incorporate randomness.  
Due to the identification (3.1), a change in {\it sign} of the Ising
exchange, corresponds to 
a sign change of a node
parameter $\theta_i$ in the network model.  Thus the random-bond
Ising model corresponds to a network model in which the {\it sign}
of the node parameters is random.  To be specific, we choose
the {\it magnitude} of the node parameters to be constants $\theta_1$
and $\theta_2$, satisfying the isotropy condition $\sinh(\theta_1)
\sinh(\theta_2)=1$.  The {\it sign} of $\theta$ at each node is chosen
randomly, being negative with probability $W$ and
positive with probability $1-W$.  The random network model 
we consider is thus
characterized by two parameters: $\Delta$, which 
measures the distance from criticality in the pure model, and $W$ the disorder
strength.

This model differs in symmetry from the original
Chalker-Coddington network model,\cite{chalker88} in which random fluxes were present through each plaquette,
reflecting the breaking of time-reversal invariance by the
magnetic field in the QHE.
In the present case, there are no random phase factors.
However, in the pure model with all
node parameters positive, the fermion amplitude picks up a minus
sign upon encircling any elementary plaquette, equivalent
to a uniform flux $\pi$ through
each plaquette.  Moreover, a sign change of a node parameter, changes
the flux through two neighboring plaquettes
by $\pi$, as depicted in Fig. 4.  

\begin{figure}
\epsfxsize = 1.6in
\centerline{ \epsfbox{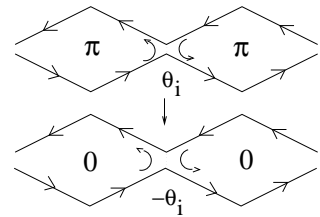}}
\begin{minipage}[t]{8.1cm}
\caption{  A sign change in a node parameter $\theta$
effectively changes by $\pi$ the flux penetrating the two
neighboring plaquettes.}
\end{minipage}
\end{figure}

Under the Ising duality transformation,
$\Delta \rightarrow
-\Delta$.  Thus the $\Delta=0$ line
should correspond to the Ising
ferromagnetic to paramagnetic phase boundary, depicted in Fig. 1.  
Increasing the randomness $W$,
with $\Delta=0$, corresponds to moving along this phase
boundary away from the pure Ising critical point.  Based on 
Fig. 1, we expect the Ising multicritical point
to correspond to some critical disorder
strength $W_c$.  Numerical simulations, described in the next section,
indeed support this scenario.  The scaling axis along the
phase boundary corresponds to varying $W$ with $\Delta=0$, whereas
the Nishimori line corresponds to the line $W=W_c$.

For $W > W_c$ and $\Delta=0$, one expects the network model to remain critical,
due to duality symmetry.  However, it is unclear what this regime corresponds to in the original lattice Ising model (1.1).  It is conceivable that
increasing $W$ beyond $W_c$ (at $\Delta=0$) corresponds to 
moving along the low-temperature part of
the phase boundary in Fig. 1, arriving at the $T=0$ fixed point
at the maximum disorder strength, $W_{\rm max} =1/2$.  However, this interpretation
is a bit problematic since $W=1/2$ is ``halfway" between
the ferromagnet and antiferromagnet, and naively corresponds
to $p=1/2$ in Fig. 1.  Perhaps the time continuum limit
taken in Eq. (2.1), does {\it not} give a faithful representation
of the lattice Ising model (1.1) for $W>W_c$.

\section{results}

The transfer matrix $T$ of the random network model can be
computed numerically, following the work of Chalker\cite{chalker88} 
and others.\cite{dhlee93,liu94,dklee94}
We have studied strips of width $L$ ranging from $L=16$ up to $L=128$.
Ensemble averaging is performed by taking very long
strips, with length $N$ up to $10^5$.  Of the $L$ eigenvalues of $T_N$,
denoted $\lambda_i$, $L/2$ are greater than 1, corresponding
to exponentially growing solutions, and the others are less than 1,
decaying to zero with increasing $N$.  Due to the (statistical) parity
of the system, these come in pairs, 
$\lambda_i = \exp(\pm N \gamma_i)$, where
$i = 1,2,...,L/2$ and all the $\gamma_i^s$ are positive.
Of interest is the smallest, $\gamma_{\rm min}$, corresponding to
the most slowly decaying mode.  From this one extracts the correlation length
as
\begin{equation}
\xi_L =  {1 \over {\gamma_{\rm min}}}.
\end{equation}

In order to extrapolate to the thermodynamic limit, $L \rightarrow \infty$,
it is convenient to consider the dimensionless ratio,
\begin{equation}
\Lambda_L(\Delta,W)  \equiv
{L \over \xi_L} ,
\end{equation}
which is a function of the two control parameters $W$ and $\Delta$.
Away from criticality ($\Delta \ne 0$), $\xi_{L=\infty}$ is finite, and so this ratio should grow and diverge
as $L \rightarrow \infty$.  
Representative data are shown in Fig. 7,
where $\Lambda_L$ is plotted versus $\Delta$ for weak
randomness $W=0.075$, at various different system sizes.  
At the pure Ising critical point
$\Delta = W = 0$, the ratio $\Lambda_L$ is found to vanish identically
even for {\it finite} $L$, due to the propagating mode
with transverse momentum $p=\pi$, described by Eq. (3.2).
At a random critical point, one expects that $\Lambda_L$
will approach a finite constant in the thermodynamic limit,
reflecting the infinite correlation length.

In Fig. 5 we show data for $\Lambda_L$ versus disorder strength
$W$, along the phase boundary, $\Delta = 0$, at four different
system widths $L$.  Although there is significant variation with
$L$, particularly for the smaller sizes, there appear
to be two distinct regimes separated by a peak.  
For the largest width ($L=128$) the peak occurs at a disorder strength
$W_c \approx 0.08$.  For $W < W_c$ the ratio $\Lambda_L$
drops rapidly towards zero, the value at the pure Ising critical point.
For strong disorder $W >W_c$ the ratio appears to be settling down
towards a constant of order 1/2 at large $L$.  This presumably corresponds
to a strong disorder critical point.  Right at the critical disorder strength
$W=W_c$, $\Lambda_L$ is increasing slowly with $L$, and presumably eventually saturates.  We thus identify the point ($\Delta=0, W=W_c$) with
the random Ising multicritical point (see Fig. 1).

\begin{figure}
\epsfxsize = 3in
\centerline{ \epsfbox{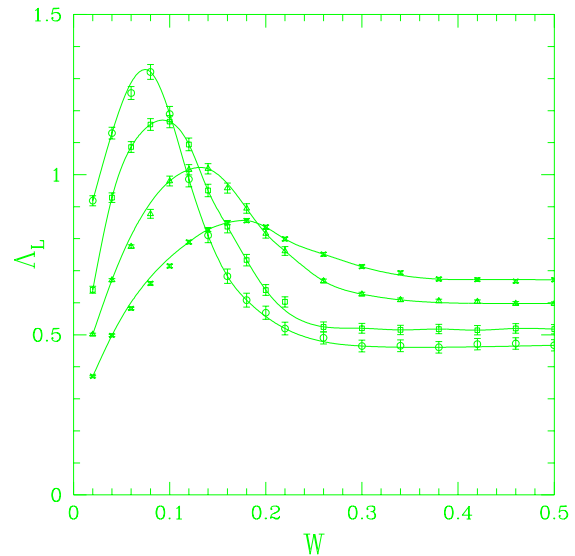}}
\begin{minipage}[t]{8.1cm}
\caption{Dimensionless ratio $\Lambda_L$
vs disorder strength $W$ along the
phase boundary $\Delta = 0$ for four different system widths:
$L=16(\times)$, $L=32(\triangle)$, $L=64(\Box)$, and $L=128(\bigcirc)$.}
\end{minipage}
\end{figure}

This identification can be confirmed by extracting critical exponents,
and comparing with the values obtained 
by Singh and Adler\cite{singh96} at the Ising
multicritical point. 
Unfortunately, 
precise values are difficult to extract due to the rather
severe finite size effects, evident in Fig. 5.  
These are due (in part) to the small value of $W_c$:  With
$W_c \approx 0.08$, the typical nearest distance between
two nodes with negative node parameters, $\theta_i$, is roughly
4 times the network lattice spacing.  

Consider first the critical exponent corresponding to the scaling
axis along the phase boundary, denoted $\nu_p$.  
A natural finite-size scaling ansatz for $\Lambda_L$
takes the form,
\begin{equation}
\Lambda_L(\Delta=0, W) = f[L^{1/\nu_p} (W-W_c)]  ,
\end{equation}
for large $L$ and $W \rightarrow W_c$.  This form predicts that
peaks in $\Lambda_L(W)$ around $W_c$ should
sharpen up with a width vanishing
as $\delta W \sim L^{-1/\nu_p}$.  The data in Fig. 5 are 
consistent with this trend, showing narrower peaks for larger
$L$.  To obtain a rough estimate for
the exponent $\nu_p$, we fit the peaks to a parabola. 
Denoting the curvatures of the parabolas as $R_L$, scaling predicts
$R_L \sim L^{2/\nu_p}$.  In Fig. 6 we plot 
$\log(R_L)$ versus $\log(L)$ and extract the exponent
$2/\nu_p$ as the slope of a fitted straight line.
Fitting all four points gives an estimate
$\nu_p \sim 2.2$.  
However, the $\chi^2$ per data points decreases an order of magnitude
if the smallest size $L=16$ is excluded, which gives (dotted line in Fig. 6)
$\nu_p\sim 2.45$.
Thus we estimate $\nu_p  \approx 2.4$, with a large error bar $\pm 0.3$.

\begin{figure}
\epsfxsize = 2.5in
\centerline{ \epsfbox{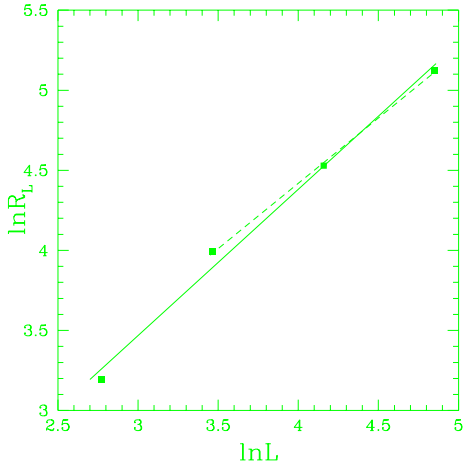}}
\begin{minipage}[t]{8.1cm}
\caption{ Curvatures $R_L$ of parabolic fits
to the peaks in Fig. 5, vs
system width $L$ on a doubly logarithmic plot.
 From scaling, the slope gives an estimate for the critical exponent
$2/\nu_p$.}
\end{minipage}
\end{figure}

The critical exponent along the Nishimori line
can be extracted by sitting at $W_c$ and tuning $\Delta$ away from zero.
In this case, finite-size scaling implies that
\begin{equation}
\Lambda_L(\Delta, W=W_c) = F[L^{1/\nu} \Delta]  .
\end{equation}
The raw data for $\Lambda_L(\Delta, W_c)$ versus $\Delta$ 
are shown in Fig. 7.  As expected, away from the multicritical point
at $\Delta=0$, the ratio $\Lambda_L$ grows with $L$, indicative
of a finite correlation length. 
In Fig. 8 these data for $\Delta < 1/2$ are replotted,
rescaling the horizontal axis by $L^{1/\nu}$.
Based on the quality of the data collapse, we estimate
$\nu \approx 4/3$
with error bars $\pm 0.1$.

Our estimates for the exponents compare favorably
with those obtained by
Singh and Adler\cite{singh96}: $\nu=1.32\pm 0.08$
and $\nu_p \approx (5/3) \nu \approx 2.2$.
This agreement gives one confidence that the
peak in Fig. 5 does indeed correspond to
the random Ising multicritical point.

In addition to the multicritical point,
we have tried to analyze the behavior at maximal
disorder strength $W=1/2$.
As discussed earlier, it is unclear whether or not this
point corresponds to the $T=0$ fixed point at $p=p_c$ in the
random-bond Ising model (Fig. 1).  Unfortunately,
at $W=1/2$ we are even more severely plagued by finite-size effects
as $\Delta$ is varied.  Specifically, the correlation length
tends to remain very long, even well
away from criticality, with $\Delta \rightarrow 1$.
In this limit, the fermions tend to become localized
around plaquettes on one sublattice.  But with
$W=1/2$, half of the plaquettes have zero flux,
and can support ($E=0$) states circling around them.
For $1 - \Delta$ small, these states will be weakly coupled
via tunneling, and may tend to percolate out to rather long scales.
At this stage, we cannot conclude anything definite about
the critical behavior of the network model at strong disorder.

\begin{figure}
\epsfxsize = 3in
\centerline{ \epsfbox{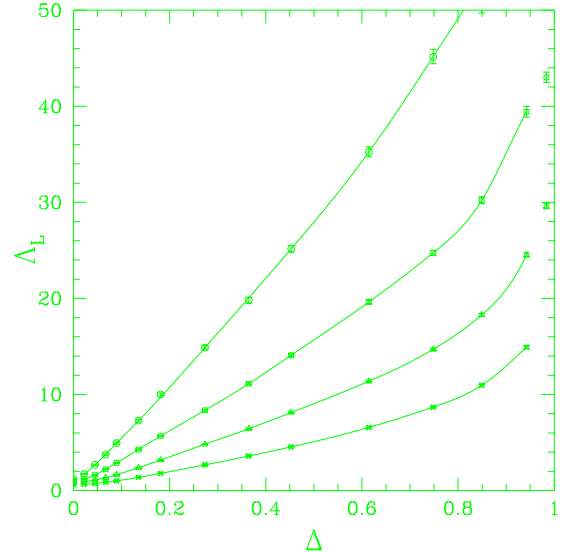}}
\begin{minipage}[t]{8.1cm}
\caption{ Dimensionless ratio $\Lambda_L=L/\xi_L$
plotted vs $\Delta$ at disorder strength
$W=0.075$ for four different system sizes:
$L=16(\times)$, $L=32(\triangle)$, $L=64(\Box)$, and $L=128(\bigcirc)$}
\end{minipage}
\end{figure}

\section{conclusion}

To summarize, we have shown that the 2D random-bond
Ising model can be fruitfully mapped onto
a network model for chiral fermions.
The network model is similar to the original 
Chalker-Coddington model\cite{chalker88}
used to study IQHE plateau transitions, but with different symmetry.
Specifically, the model has node
parameters with random signs, rather than random fluxes
through plaquettes.  
The network model has been used to study the novel
random multicritical point which exists on the
Ising ferromagnetic to paramagnetic phase boundary.  
By implementing a numerical transfer matrix approach,
estimates for the associated critical exponents
have been extracted, which are consistent with those obtained
by Singh and Adler\cite{singh96} from high-temperature series.
The critical exponents are quite different
from those at the IQHE plateau transition,
indicating different universality classes, not surprising
in view of the symmetry differences between the two models.

\begin{figure}
\epsfxsize = 3in
\centerline{ \epsfbox{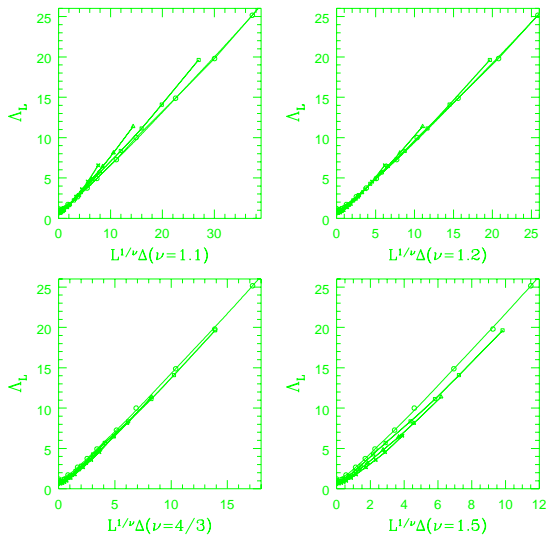}}
\begin{minipage}[t]{8.1cm}
\caption{ Scaling collapse of the data from Fig. 7
with $\Delta <1/2$
for four different exponent values: $\nu=1.1$, $1.2$, ${4\over 3}$,
and $1.5$.}
\end{minipage}
\end{figure}

What are the prospects for an analytic treatment of the
random Ising multicritical point?  Being in two dimensions,
one might hope that powerful constraints from conformal
invariance would be helpful.  Analytic approaches to the
IQHE plateau transition have been impeded by the absence of
critical behavior in ensemble-averaged single-particle
Green's functions.  The situation might be simpler
at the Ising multicritical point, though, since critical
properties are probably present in average single-fermion
correlators.

\section*{Acknowledgement}

It is a pleasure to acknowledge fruitful conversations with
Leon Balents, Daniel Fisher, Derek Lee, Dongzi Liu,  and Andreas Ludwig.  
We are grateful to the National Science Foundation for support, under 
Grant Nos. PHY94-07194, DMR-9400142, and DMR-9528578.  

\bibliography{/usr/home/spock/sora/paper/biblio/ref}
\end{multicols}
\end{document}